\documentclass[aps,twocolumn,notitlepage,superscriptaddress,showkeys,showpacs]{revtex4-1}
\usepackage{amsmath,amsfonts,amssymb,amsthm,epsfig,array,physics}
\usepackage{graphicx,tabularx}
\usepackage{upgreek}
\usepackage[etex=true,export]{adjustbox}
\usepackage{dsfont}
\usepackage{graphics}
\usepackage{float}
\usepackage{bm}
\usepackage{verbatim}
\usepackage{gensymb}
\usepackage[dvipsnames]{xcolor}
\usepackage[hidelinks,colorlinks,linkcolor=blue,
citecolor=blue,urlcolor=blue]{hyperref}
\usepackage[titletoc,title]{appendix}

\newcommand{\beq}{\begin{equation}}
\newcommand{\eeq}{\end{equation}}
\newcommand{\beql}{\begin{equation*}}
\newcommand{\eeql}{\end{equation*}}
\newcommand{\beqn}{\begin{eqnarray}}
\newcommand{\eeqn}{\end{eqnarray}}

\renewcommand{\vec}[1]{\mbox{\boldmath$#1$}}

\begin{document}

\title{ Electrostatic environment and Majorana bound states in full-shell topological insulator nanowires  }

\author{Li Chen}
\affiliation{School of Physics, Huazhong University of Science and Technology, Wuhan, Hubei 430074, China}
\affiliation{State Key Laboratory of Low Dimensional Quantum Physics, Department of Physics, Tsinghua University, Beijing, 100084, China}

\author{Xiao-Hong Pan}
\affiliation{School of Physics, Huazhong University of Science and Technology, Wuhan, Hubei 430074, China}

\author{Zhan Cao}
\affiliation{Beijing Academy of Quantum Information Sciences, Beijing 100193, China}

\author{Dong E. Liu}
\affiliation{State Key Laboratory of Low Dimensional Quantum Physics, Department of Physics, Tsinghua University, Beijing, 100084, China}
\affiliation{Frontier Science Center for Quantum Information, Beijing 100184, China}
\affiliation{Beijing Academy of Quantum Information Sciences, Beijing 100193, China}

\author{Xin Liu}
\email{phyliuxin@hust.edu.cn}
\affiliation{School of Physics, Huazhong University of Science and Technology, Wuhan, Hubei 430074, China}

\date{\today}

\begin{abstract}

The combination of a superconductor (SC) and a topological insulator (TI) nanowire was proposed as a potential candidate for realizing Majorana zero modes (MZMs). In this study, we adopt the Schr\"odinger-Poisson formalism to incorporate the electrostatic environment inside the nanowire and systematically explore its topological properties. Our calculations reveal that the proximity to the SC induces a band bending effect, leading to a non-uniform potential across the TI nanowire. As a consequence, there is an upward shift of the Fermi level within the conduction band. This gives rise to the coexistence of surface and bulk states, localized in an accumulation layer adjacent to the TI-SC interface. When magnetic flux is applied, these occupied states have different
flux-penetration areas, suppressing the superconducting gap. However, this impact can be mitigated by increasing the radius of the nanowire.
Finally, We demonstrate that MZMs can be achieved across a wide range of parameters centered around one applied flux quantum, $\phi_0 = h/2e$. 
Within this regime, MZMs can be realized even in the presence of conduction bands, which are not affected by the band bending effect.
These findings provide valuable insights into the practical realization of MZMs in TI nanowire-based devices, especially in the presence of a complicated electrostatic environment.


\end{abstract}

\maketitle

\section{Introduction}

Majorana zero modes (MZMs), as quasi-particles at topological
superconductor boundaries, have been extensively studied because of their potential applications in topological quantum computations~\cite{KITAEV20032,Nayak-RevModPhys-2008,RevModPhysQI2011}. 
The most heavily
investigated experimental systems to search for MZMs are semiconductor (SM)-superconductor (SC) devices~\cite{Lutchyn2010,Oreg2010,Sau2010}. Despite various experimental progress been reported~\cite{Mourik2012,Deng2012,Rokhinson2012_np,Wiedenmann2016_nc,cao2023recent}, 
the conclusive observation of MZMs is still lacking. A significant reason is that some trivial mechanisms can also produce similar experimental signatures~\cite{Kells-prb-2012,Lee-prl-2012,Liu-prb-2017,Moore-prb-2018,Vuik-scip-2019,Chen-prl-2019,Awoga-prl-2019,junger-prl-2020,Prada-np-2020,Valentini-science-2021}, which significantly complicated the search for MZMs. 
To overcome this issue, two main directions have been pursued. The
 first approach involves utilizing alternative detection methods providing signals that can hardly be mimicked by non-Majorana states~\cite{Szumniak-prb-2017,Chevallier-prb-2018,Fleckenstein-prb-2018,Chen-prb-2019,Legg-prb-2022}. One such method is nonlocal conductance measurements in three-terminal devices~\cite{Gramich-prb-2017,Rosdahl-prb-2018,Danon-prl-2020,Menard-prl-2020,Pan-prb-2021,Pikulin2021,Wang-prb-2022,Poeschl2022-PRB,Banerjee-prl-2023}, which can directly detect the bulk gap closing and
reopening. The second approach focuses on finding materials with high quality and unique properties that are conducive to the formation and manipulation of MZMs~\cite{Cook-prb-2011,Cook-prb-2012,Legg-prb-2021,Schellingerhout2021,Cao-prb-2022,Jiang-prm-2022,Geng-prb-2022,Kate-nanole-2022,pan2023meissner}. For instance, materials like topological insulator (TI) nanowires have been identified as potential candidates~\cite{Cook-prb-2011,Cook-prb-2012,Legg-prb-2021}. When a TI is made into a nanowire, quantum confinement
gives rise to peculiar one-dimensional Dirac sub-bands whose energy dispersion can
be manipulated by external fields.
In contrast to semiconductor-based systems where the Fermi level needs to be finely tuned within a narrow gap opened by the Zeeman effect, TI nanowires offer a topological region that can extend throughout the entire bulk gap~\cite{Cook-prb-2011,Cook-prb-2012}.

In the past few years, substantial progress in
the growth of TI nanowire devices have been reported~\cite{Peng-NM-2010,Zhang-prl-2010,Xiu-NT-2011,Cho-NC-2015}. These advancements enabled the fabrication of high-quality TI nanowires with well-controlled properties~\cite{Muenning-NC-2021a,Legg-NN-2022a,Roessler-NL-2023}. 
Recently, proximity-induced superconductivity in TI nanowires have been experimentally reported~\cite{Fischer-prr-2022,Bai_2022, Roessler-NL-2023}, but the deterministic evidence of the MZMs in TI nanowire is still lacking. 
Meanwhile, the previous theoretical works~\cite{Cook-prb-2011,Cook-prb-2012,Legg-prb-2021} treat chemical potential and induced superconducting (SC) gap as
 as independently adjustable parameters.
  However, the unavoidable electrostatic effects and band bending effect at the TI-SC interface can greatly complicate the Majorana physics in the  TI-SC system, as they did in the SM-SC system~\cite{Vuik-IOP-2016,Antipov2018,Mikkelsen2018}.
 For instance, in experiments where TI are grown with SC films, the process induces charge doping from the SC to the TI, resulting in a shift of the Fermi level into the conduction band~\cite{Xu2014,Xu-prl-2015}.
 This effect is undesirable since the realization of MZMs requires TIs to be bulk-insulating~\cite{Hosur2011,Cook-prb-2011,Cook-prb-2012}. Furthermore, the band bending effect near the SC can suppress the tunability of surface states through gating~\cite{Chen-prb-2023}, further complicating the control of electronic properties in TI-SC hybrid devices. 
These challenges and limitations motivate us to develop more realistic calculations that can accurately describe the electrostatic environment and band structures of TI nanowires, leading to a better understanding of their topological properties.

In this work, we investigate the properties of a TI nanowire covered by a full-shell SC. To account for the electrostatic environment of the system, we employ the self-consistent Schr\"odinger-Poisson (SP) methods to calculate the electrostatic potential inside the TI nanowire.
Our analysis reveals that the band bending effect at the TI-SC interface leads to a shift of the Fermi level into the conduction band, consistent with experimental observations~\cite{Xu2014,Xu-prl-2015}.
Consequently, the surface states and bulk states coexist in the system, and they are confined to an accumulation region near the TI-SC interface.
Moreover, these occupied states have different flux-penetration areas, leading to a suppression of the SC gap
under the application of a magnetic field.  
To address this issue, we propose to use a TI nanowire with a larger radius.
Finally, we give a topological phase diagram and demonstrate that MZMs can be achieved over a wide range of parameters near one applied flux quantum, $\phi_0 = h/2e$. In this case, the presence of MZMs is independent of the strength of the band bending, eliminating the need for fine tuning of the Fermi level. These findings provide valuable insights into the phase diagram and practical realization of MZMs in TI nanowire-based devices.

The paper is organized as follows. In Sec.~\ref{sec_mode}, we construct a model Hamiltonian in the cylinder coordinate.  
In Sec.~\ref{sec_potential}, we calculate the electrostatic
potential using the Schr\"{o}dinger-Poisson approach. In Sec.~\ref{sec_tsc}, we discuss the topological properties of the TI nanowire.
Finally, we draw a discussion and conclusion in
Sec.~\ref{sec_conclusion}.

\section{Model Hamiltonian}\label{sec_mode}

\begin{figure}[!htb]
\centerline{\includegraphics[width=1\columnwidth]{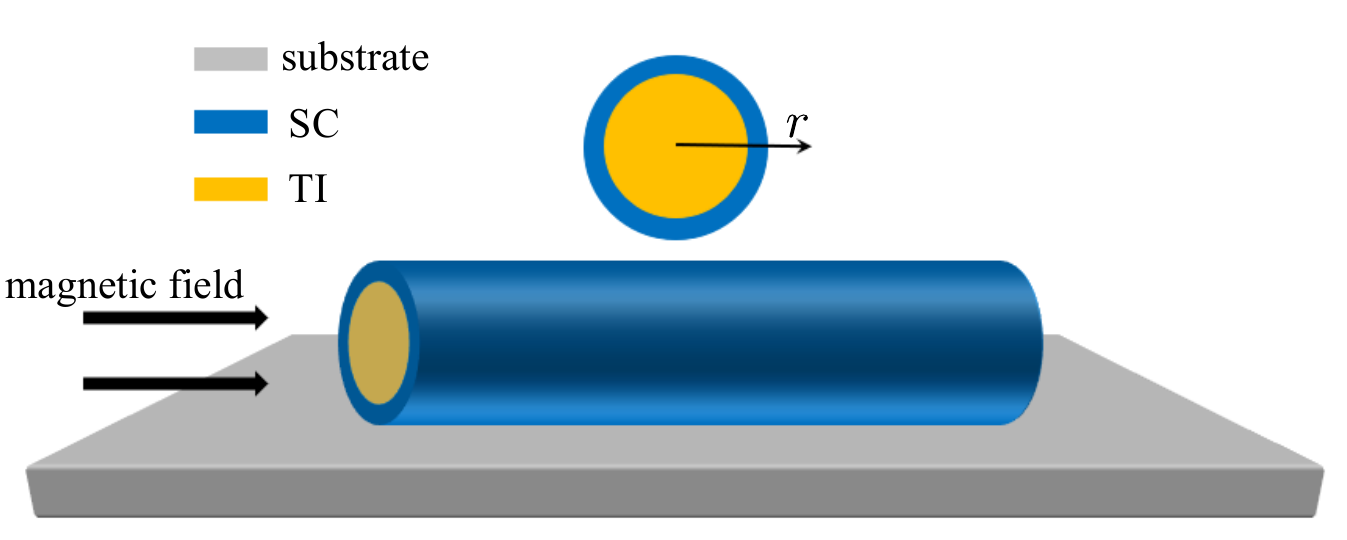}}
\caption{ A TI nanowire is covered by a full superconducting shell. The magnetic field is applied along the nanowire ($z$ direction). The radius of the TI nanowire is $R_0$.}
 \label{set_up}
\end{figure}

We consider a topological insulator (TI) nanowire coated by a full superconducting shell, as illustrated in Figure~\ref{set_up}. The system is exposed to a magnetic field $\vec{B}$ oriented along the nanowire's direction. To maintain the system's rotational symmetry, we adopt the electromagnetic vector potential $\vec{A} = \frac{1}{2}(\vec{B} \times \vec{r})$. Subsequently, we formulate the electronic Hamiltonian of the TI in cylindrical coordinates as follows (see Appendix~\ref{Appendix A}):
\beqn\label{Ham_e}
H_{\rm e} = H_{\rm TI} + H_{M}- e\phi(r),
\eeqn
where
\begin{eqnarray}\label{Ham1}
 H_{\textrm{TI}}& = & M(r,\theta,z) s_0\sigma_z+D(r,\theta,z) s_0\sigma_0 +A_1 (-i\partial_z)s_z\sigma_x \nonumber  \\ 
&& +A_2 P_{-\theta}s_{+}\sigma_x +A_2 P_{+\theta}s_{-}\sigma_x.
\end{eqnarray}
The Pauli matrices $s$ and $\sigma$ acts on spin and orbital space, respectively. 
$r$, $\theta$, and $z$ are the cylindrical coordinates.
We define $s_{\pm} =  (s_x \pm i s_y)/2$, $s_{\theta}=\cos\theta s_y - \sin\theta s_x$,  $M(r,\theta,z) = m_0 - B_1 \partial_{z}^{2}-B_2(\frac{1}{r}\partial_r +\partial_{r}^{2}+
\frac{1}{r^2}\partial_{\theta}^2)$, $D(r,\theta,z) = C_0 - D_1 \partial_{z}^{2}-D_2(\frac{1}{r}\partial_r +\partial_{r}^{2}+
\frac{1}{r^2}\partial_{\theta}^2)$, $P_{\pm \theta} = -i e^{\pm i \theta}(\partial_r \pm \frac{i}{r}\partial_{\theta})$. 
The parameters $m_0,~C_0,~B_i,~A_i$ and $D_i$ with $i = 1,2$ are model parameters from \emph{ab initio} calculations~\cite{Zhang2009_NP}. 
The electrostatic potential $\phi(r)$ arises due to the band bending effect at the interface between the TI and SC, which can be self-consistently calculated through the SP method, as we will show in Sec.~\ref{sec_potential}.
$H_M$ is the magnetic flux-induced term, which takes the form (see Appendix~\ref{Appendix A}):
\beqn
H_{M} = \frac{B_2}{r^2}[\Phi^2(r)+2iB_2\Phi(r)\partial_\theta]s_0\sigma_z-\frac{A_2\Phi(r)}{r}s_{\theta}\sigma_x.
\eeqn
Here, $\Phi(r)=Br^2/\Phi_0$ represents the normalized magnetic flux with respect to the flux quantum $\Phi_0 =h/e$.

\begin{figure}[b]
\centerline{\includegraphics[width=1\columnwidth]{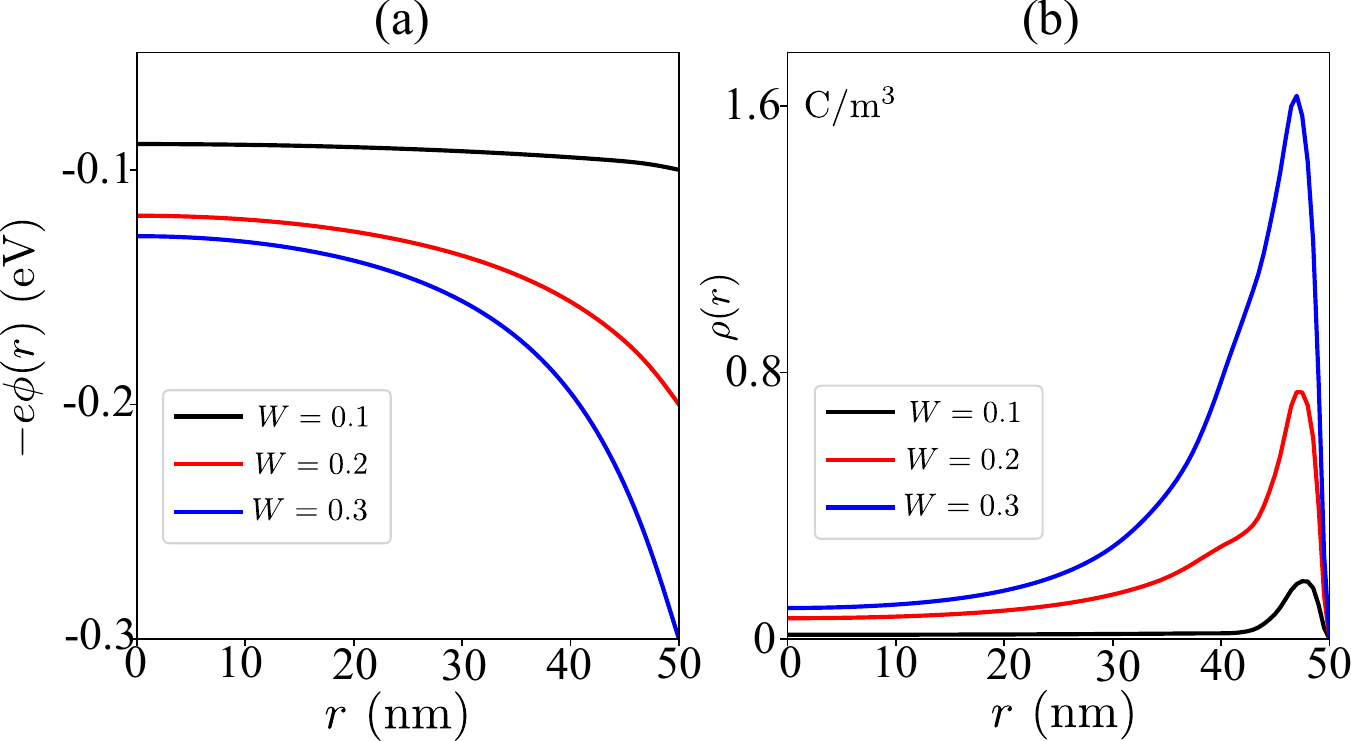}}
\caption{The profiles of (a) the electrostatic potential $-e\phi(r)$ and (b) the charge density $\rho(r)$ are depicted along the radial direction of the nanowire for three distinct band bending strengths $W$.  }
\label{potential}
\end{figure}

 \begin{figure*}[tb]
\centerline{\includegraphics[width=1.6\columnwidth]{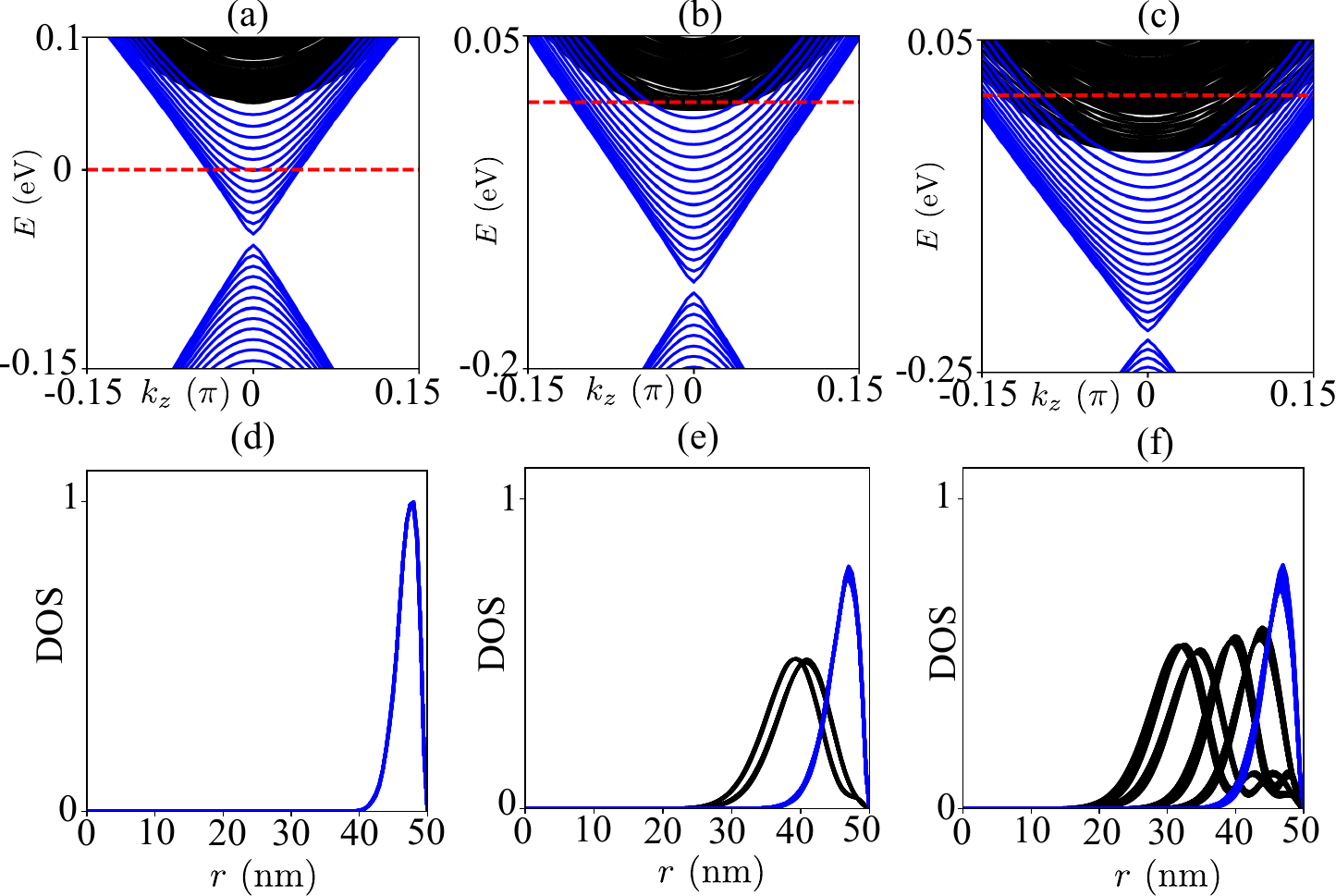}}
\caption{(a)-(c) Left to right: the band structure of TI nanowire with inhomogeneous potential $\phi(r)$ when $W= 0.1, 0.2, 0.3$ eV, respectively. The blue (black) lines correspond to surface states (bulk states). The red dashed line represents the Fermi level. Notably, the absence of a magnetic flux maintains the doubly degenerate nature of all bands due to time reversal symmetry. (d)-(e)  The density distribution of occupied states at the Fermi level in panels (a)-(c). } 
\label{potential-band}
\end{figure*}

Notably, the angular momentum operator $\hat{J}_{z}^{\rm e}$ commutes with $H_{\rm e}$, with $ \hat{J}_{ z}^{\rm e} = -i \partial_{\theta}+\frac{1}{2}s_z$. This leads to the eigenvalue $j_{\rm e}$ of $\hat{J}_{ z}^{\rm e}$ taking half-integer values $\mathbb{Z}+\frac{1}{2}$. The angular dependence of $H_e$ can be eliminated using a unitary transformation $U = \textrm{exp}[-i(j_e-\frac{1}{2}s_z)\theta]$, namely $\tilde{H}_{e} = UH_{e}U^{\dagger}$. Consequently, $\tilde{H}_{e}$ becomes block diagonal, expressed as:
\beqn\label{block-1}
&&\tilde{H}_{\rm e} = \bigoplus_{j_e,k_z} H_{\rm TI}^{j_{\rm e}}(r,k_z).
\eeqn 
Notably, we replace $-i\partial_z$ with $k_z$ because it is a good quantum number for an infinite nanowire. The explicit form of $H_{\rm TI}^{j_{\rm e}}(r,k_z)$ is given in Appendix~\ref{Appendix A}.  When the electrostatic potential is absent, i.e., $\phi(r)=0$, the energy spectrum of the surface states can be approximated by the formula~\cite{Rosenberg-PRB-2010,Ostrovsky-PRL-2010}:
\begin{equation}\label{eq-ss}
E_{k_z,j_e} = A_2\sqrt{k_z^2+\left(\frac{j_e-\Phi(R_0)}{R_0}\right)^2}.    
\end{equation}
In the absence of a magnetic field, the branches $E_{k_z,\pm |j_e|}$ are doubly degenerate due to time reversal symmetry. Upon application of a magnetic field, a finite gap of $2\delta = \frac{2 A_2\Phi(R_0)}{R_0}$ emerges between bands with $\pm j_e$. 
For $R_0 = 50$ nm, the surface level spacing is
8.2 meV at half flux quantum, i.e., $\Phi(R_0) = 1/2$. 
The corresponding Zeeman energy scale $E_z = 0.056$ meV (taking $g$ factor for Bi$_2$Se$_3$ is $ g \approx 4$~\cite{Liu-PRB-2010}). Therefore, the Zeeman effect is negligible compared to orbital effects in our system.

\section{Electrostatic Potential}\label{sec_potential}
To compute the electrostatic potential $\phi(r)$ self-consistently, we begin by solving the Schr\"odinger equation:
\begin{eqnarray}\label{sp1}
H_{\textrm{TI}}^{j_{\rm e}}(r,k_z)\psi_{n_z,k_z}^{j_e}(r) = E_{n_z,k_z}^{j_e}\psi_{n_z,k_z}^{j_e}(r)
 \end{eqnarray}
in each $j_{\rm e}$ block with given $k_z$. We solve it numerically on the basis of Bessel functions (see Appendix.~\ref{Appendix B}). Here, $n_z$ is the index of the transverse modes. 
It is important to note that we solve the Schr\"odinger equation only within the TI region. This is due to the fact that the superconductor screens the electric field due to its metallic nature. As a result, throughout the self-consistent procedure, we treat the SC shell solely as a boundary condition with a band offset $W$ at the TI-SC interface.
Then the charge density with the profile $\phi(r)$ is obtained 
\begin{equation}\label{SP-charge}
    \rho(r) = \frac{-e}{(2\pi)^2}\sum_{ n_z,j_e}\int dk_z \left[| \psi_{n_z,k_z}^{j_e}(r) |^2 f_T-\rho_{\textrm{val}}(r)\right], 
\end{equation}
where $f_T = 1/\left(e^{E_{n_z,k_z}^{j_e}/T}+1\right)$ represents the Fermi distribution. Notably, the first term on the right-hand side of Eq.~\eqref{SP-charge} accounts for the charge density originating from all occupied states. To obtain the charge density of free electrons or holes, the density from the entire valence band $\rho_{\textrm{val}}(r)$ needs to be subtracted~\cite{Chen-prb-2023}, see details in Appendix~\ref{Appendix C}. 
Finally, the electrostatic potential is determined by solving the Poisson equation in radial coordinates:
\begin{eqnarray}\label{sp2}
\frac{1}{r}\partial_r \phi(r) + \partial_r^2 \phi(r) =  - \frac{\rho(r)}{\epsilon_0 \epsilon_r}, 
\end{eqnarray}
where $\epsilon_r $ is the relative dielectric constant of TI.
The SP method is to solve Eq.\eqref{sp1} and Eq.\eqref{sp2} self-consistently (see Appendix~\ref{Appendix D} for details).

In Figure~\ref{potential}, we present the distribution of the self-consistent potential $\phi(r)$ and charge density $\rho(r)$ for various values of $W$. Notably, the potential gradually increases from the boundary to the interior of the TI due to the charge screening effect. The contact between the TI nanowire and the SC shell induces charge doping from the SC to the TI, resulting in an upward shift of the Fermi level, which is evident in the band structure shown in Figure~\ref{potential-band}(a)-(c). When $W$ is relatively small, the Fermi level remains within the bulk band gap, leading to the occupation of only surface states [Figure~\ref{potential-band}(a)]. As $W$ increases up to 0.2 and 0.3 eV, the Fermi level moves into the conduction band [Figure~\ref{potential-band}(b)-(c)]. Recent \emph{ab initio} calculations suggest that $W \approx 0.3$ eV in Bi$_2$Te$_3$-Nb hybrid systems~\cite{Ruessmann2022}. This implies that most TI-SC nanowires naturally exhibit a Fermi level pinned within the conduction band, as demonstrated in Figure~\ref{potential-band}(c).
Furthermore, the distributions of the density of states (DOS) of the wave functions at the Fermi level are shown in Figure~\ref{potential-band}(d)-(f). Remarkably, both the occupied surface states and bulk states are confined to a narrow accumulation region near the TI-SC interface~\cite{Bahramy2012-nc,Michiardi2022-nc}, characterized by a width of about 30 nm. The remaining part of the nanowire remains relatively insulating in the bulk. Moreover, it is observed that the surface states and bulk states exhibit distinct localizations near the TI-SC interface, as indicated by the blue and black lines in Figure~\ref{potential}(e)(f). This confinement of the surface states and bulk states within the accumulation region is a significant consequence arising from the electrostatic environment of the system, which has not been considered in previous theoretic works~\cite{Cook-prb-2011,Cook-prb-2012,Legg-prb-2021,Legg-prb-2021}.

\begin{figure}[t]
\centerline{\includegraphics[width=1\columnwidth]{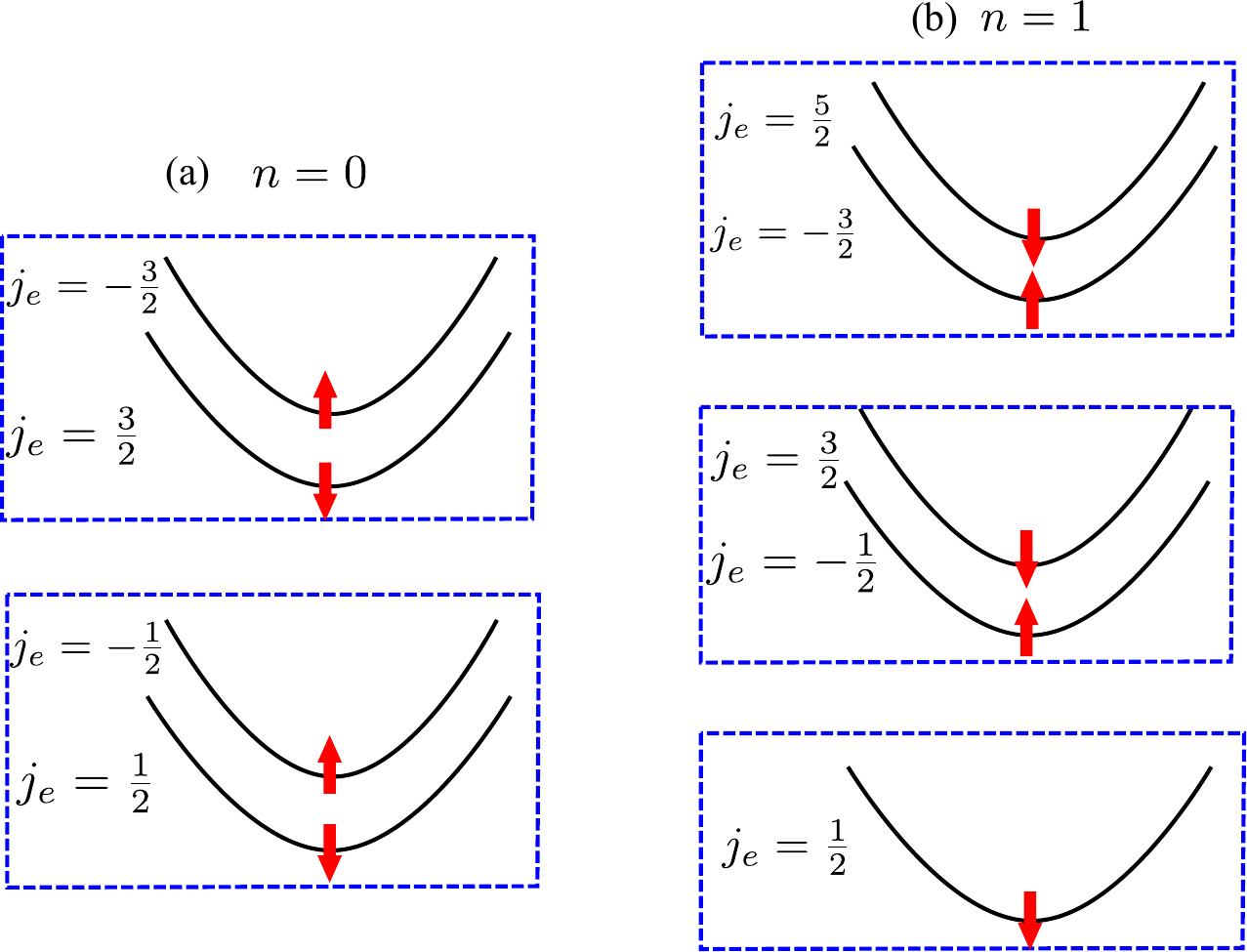}}
\caption{Schematic of the superconducting pairing sectors of the surface states for $n=0$ in panel (a) and $n=1$ in panel (b). The pairing potential occurs between two surface states whose total angular momentum satisfies $j_{\rm e1} + j_{\rm e2} = n$, as indicated by the blue dashed box. The red upward (downward) arrows signify surface states with negative (positive) angular momentum $j_e$.  }
 \label{fig-paring}
\end{figure}

\section{Topological property}\label{sec_tsc}

When considering the presence of the superconductor, the system is described by the Bogoliubov-de Gennes (BdG) Hamiltonian, which takes the form:
	\beqn \label{eq:FeSC-Hmtn}
		&& H= \begin{pmatrix}
			H_{\rm e} & is_y\Delta(r) e^{in\theta} \\
			-is_y\Delta(r) e^{-in\theta} & -H_{\rm e}^*
		\end{pmatrix}. 
	\eeqn 
We use a spatial dependence of the pairing amplitude in
such a setup, which is given by
 $\Delta(r \le R_0) = \Delta_0\exp((r-R_0)/\xi)$~\cite{Floetotto2018}, $\xi$ is superconducting coherence length in the TI. $n$ is the superconducting phase winding number. In this work, we choose $n = [\phi_{\rm flux}+0.5]$ where the square brackets indicate taking the closest integer smaller than it.  $\phi_{\rm flux} = BR_0^2/\phi_0$ representing the penetrated magnetic flux normalized by the superconducting flux quanta, $\phi_0 =  h/2e$.
The BdG Hamiltonian $H$ satisfies $[\hat{J}_z, H] = 0$ with $\hat{J}_z  = -i \partial_{\theta}+\frac{1}{2}s_z \tau_z - \frac{n}{2}\tau_z$. And $j$ is the eigenvalue of the total angular momentum $\hat{J}_z$. 
Consequently, the BdG Hamiltonian can be block diagonal as:
\beqn\label{H_j}
H=\bigoplus_{j,k_z} H^{j}(r,k_z),
\eeqn
with 
\beqn\label{e-p}
H^{j} = \begin{pmatrix}
			H^{j_{e}}_{\rm TI} & is_y\Delta(r) \\
			-is_y\Delta(r) & -(H^{n-j_{e}}_{\rm TI})^*
		\end{pmatrix}.
\eeqn
In Fig.~\ref{fig-paring}, a schematic representation of the superconducting pairing sectors for the surface states is provided, characterized by the electronic angular momentum $j_e$. Notably, the pairing potential occurs between the two surface states whose total angular momentum satisfies $j_{\rm e1} + j_{\rm e2} = n$, as indicated by the blue dashed box.

\begin{figure}[!htb]
\centerline{\includegraphics[width=1\columnwidth]{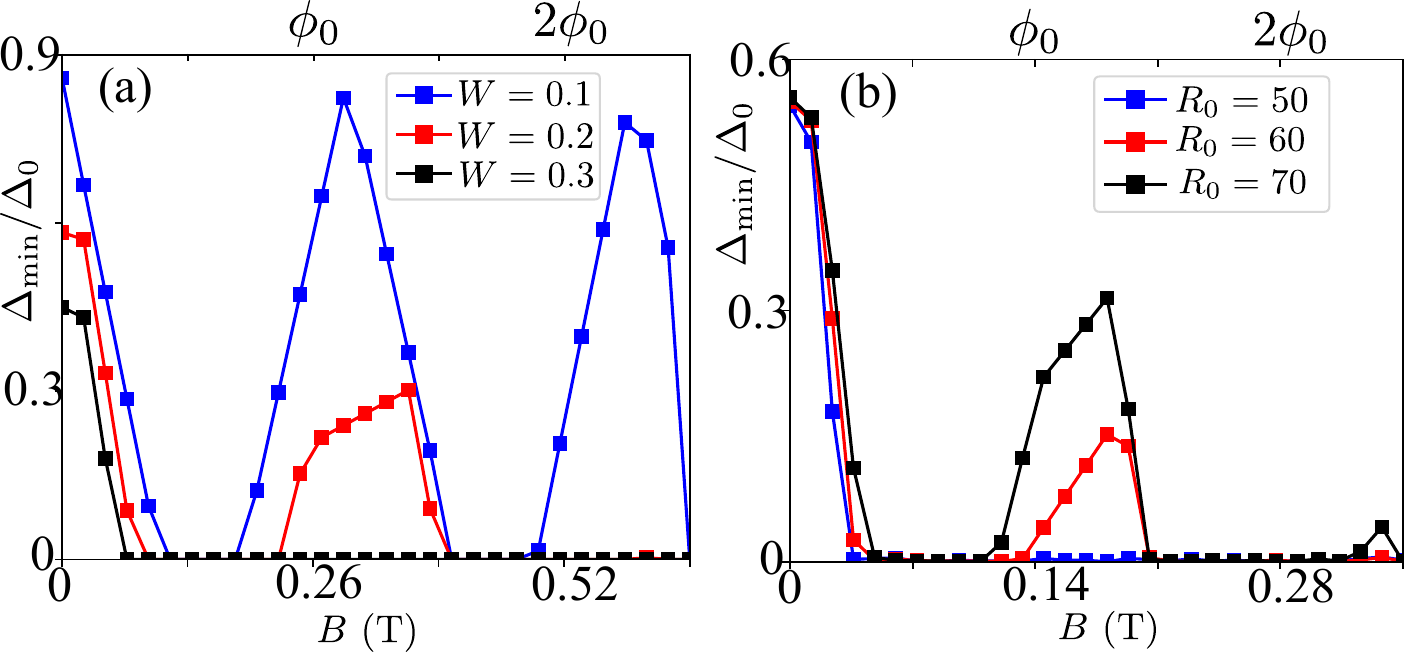}}
\caption{The minimal gap $\Delta_{\rm min}$ of all the occupied states as a function of the magnetic field with (a) different band bend strength $W$ and (b) different radius $R_0$. In panel (a),  $R_0$ is fixed to 50 nm. In panel (b), the band bending is fixed to 0.3 eV. The abscissa below panel (b) corresponds to the case with $R_0 = 70$ nm.  We choose the parameters $\Delta_0 = 1.6$ meV~\cite{Clayman-1972} and $\xi = 25$ nm~\cite{Floetotto2018}. }
 \label{fig_gap_flux}
\end{figure}

As illustrated in Fig.~\ref{potential-band}(f), the occupied surface states and bulk states exhibit distinct localizations near the TI-SC interface. Consequently, these states exhibit different magnitudes of the induced superconducting gaps. To quantitatively assess this phenomenon, we define the minimum gap $\Delta_{\rm min}$ among all occupied states.
In Fig.~\ref{fig_gap_flux}(a), $\Delta_{\rm min}$ is plotted as a function of the magnetic field for various band bending strengths $W$. When $W = 0.1$ eV (blue lines), $\Delta_{\rm min}$ is largest, displaying a typical Little-Parks oscillation behavior. Notably, the maximum of $\Delta_{\rm min}$ occurs when the flux $\phi$ slightly exceeds the integer superconducting flux quantum. This is due to the fact that the actual flux-penetration area of the states is slightly smaller than the nanowire's cross-sectional area. Furthermore, we observe a significant decrease in $\Delta_{\rm min}$ as $W$ increases, as depicted by the red and black lines in Fig.\ref{fig_gap_flux}(a). This behavior can be elucidated as follows: As $W$ rises to 0.2 eV, both surface states and bulk states become occupied [Fig.\ref{potential-band}(b)].   In general, the SC gap of bulk states is smaller than that of the surface states~\cite{Chen-prb-2023}. Additionally, their difference in the flux-penetration area, $\Delta A_{\rm phys}$, introduces a phase uncertainty $ \delta \phi = 2\pi (\Delta A_{\rm phys} B/\phi_0)$, which suppress the
$\Delta_{\rm min}$ as the magnetic field increases. 
As shown by the red (black) lines in Fig.~\ref{fig_gap_flux}(a), the third (second) Little-Parks oscillation peak disappears when $W = 0.2~(0.3)$ eV. Thus, in comparison to the scenario where the TI nanowire is solely occupied by surface states, a significant reduction in $\Delta_{\rm min}$ is observed when the Fermi level resides within the bulk bands.
To address this challenge, we propose employing a TI nanowire with a larger radius.
As illustrated in Fig.~\ref{fig_gap_flux}(b), the SC gap shows an upward trend with an increase in the nanowire's radius.
This trend
appears to be unexpected in the context of an intuitive
understanding of the proximity effect in TI-SC slab systems, in which the induced gap in the TI typically decreases with increasing thickness~\cite{Wang-science-2012,Xu2014}.
However, there are two key factors at play. Firstly, 
the presence of a full SC shell confines the occupied states to an accumulation layer near the TI-SC interface. The thickness of the accumulation layer determines the coupling strength between the TI and the SC. Notably, this accumulation layer maintains a nearly consistent thickness of approximately 30 nm, regardless of the specific radius of the TI nanowire (see Appendix~\ref{Appendix E}).  
Secondly, as the radius of the nanowire's cross-sectional area increases, the ratio between the accumulation layer and the nanowire's sectional area can be effectively reduced. As a consequence, this leads to an enhancement of $\Delta_{\rm min}$.

 \begin{figure*}[!htb]
\centerline{\includegraphics[width=1.8\columnwidth]{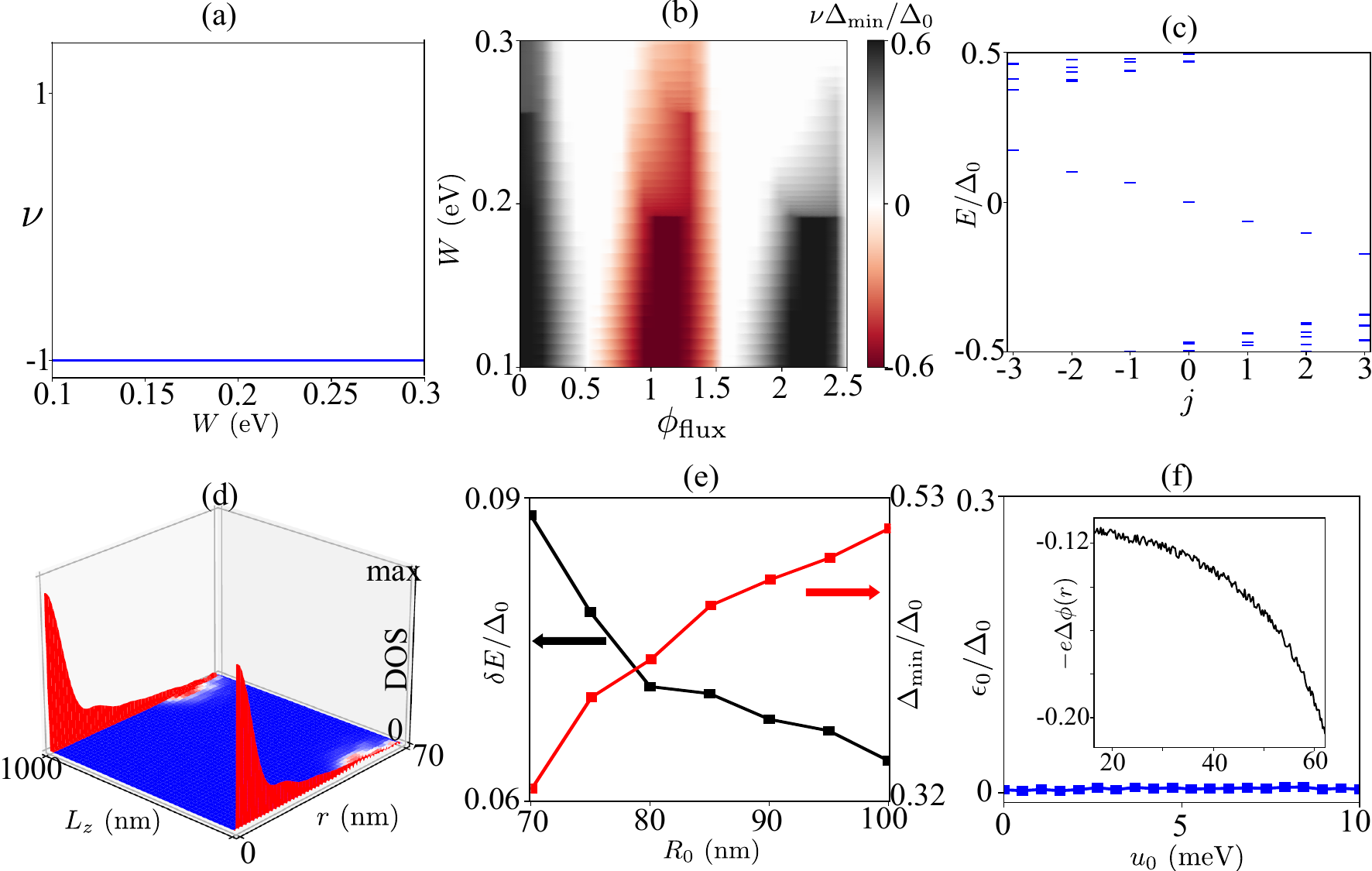}}
\caption{ (a) The topological invariant $\nu$ as a function of the band bending strength $W$ when $n=1$.  (b) The phase diagram as a function of magnetic flux and band bending strength. The SC gap is multiplied by the topological invariant $\nu$, so the red regions correspond to the gaped topological phase. (c) The eigenvalues of several lowest $j$ blocks when $n=1$. A pair of MZMs exists in $j=0$ block. (d) The distribution of DOS of MZMs in the $L_z-r$ plane. (e) Black line: the average energy separation of the in-gap states, $\delta E$ decreases with increasing $R_0$. Red line: the minimal gap $\Delta_{\rm min}$ increases with $R_0$.  (f) The average
energy of MZMs with 30
different disorder configurations at various fluctuation strengths $u_0$. The inset shows the distribution of electrostatic potential with fluctuations $ u_0 = 5$ meV in one disorder configuration. 
Parameters used in each panel: (a) $R_0 = 70$ nm, $\phi = 1.26$. (b) $R_0 = 70$ nm. (c)(d)(f) $R_0 = 70$ nm, $L_z = 1000$ nm, $\phi = 1.26$, $W = 0.3$ eV. (e) $\phi = 1.26$, $W = 0.3$ eV.  }
\label{phase_diagram}
\end{figure*}

To characterize the topology of the TI nanowire, we calculate the  the Pfaffian topological invariant $\nu$, also called the Kitaev or Majorana number~\cite{Kitaev2001}. A unitary
transformation is used to express the Hamiltonian $H$ in the Majorana basis $H_{M_j}$, which is also block diagonal as
\beqn\label{Ham-n=0}
H_{ Mj}(k_z=0,\pi) &=& \bigoplus_{j} H^{j}_{Mj}(k_z=0,\pi).
\eeqn
Then the topological invariant $\nu$ can be calculated in each $j$ blocks, which takes the form~\cite{Fernando-sci-2019}
\beqn\label{pfaeq1}
\nu &=& {\rm{sgn}}\bigg\{ \prod_{j} \frac{{\text{Pf}}[H^{j}_{Mj}(k_z = 0)]}{\text{Pf}[H^{j}_{ Mj}(k_z = \pi)]} \bigg\}\nonumber\\
&=& {\rm{sgn}}\bigg\{\prod_{j }\nu_{j} \bigg\}.
\eeqn


As depicted in Fig.~\ref{fig-paring}, the configuration of the superconducting pairing depends on the parity of the winding number $n$. This feature engenders different topological properties of the TI nanowire, contingent on whether $n$ is an even or odd integer.
For the sake of clarity, let us first consider the even-$n$ scenario, as illustrated in Fig.~\ref{fig-paring}(a).
The surface states with $\pm j_e$ exhibit an energy splitting which plays a similar role to the Zeeman splitting in the Rashba nanowire system. 
 Therefore, the realization of MZMs requires the fine tuning of the Fermi level. However, since the TI nanowire is fully surrounded by the SC shell, the strong screening effect in the SC shell makes it difficult to tune the Fermi level by the gate voltage. 
Although Fermi level control can be equivalently achieved by altering the magnitude of $W$, it is essential to note that in practical experiments, $W$ is a nonadjustable parameter that is determined by work function imbalance at the TI-SC interface~\cite{KIEJNA-1996}.
Considering these intricate factors, the realization of MZMs with even-$n$ appears to be difficult in our proposed framework.

In the scenario where $n$ is an odd integer, such as the case of $n=1$, distinct behavior emerges. Here, the presence of a solitary $j_e=1/2$ surface subband [Fig.~\ref{fig-paring}(b)] violates the Fermion doubling theorem, leading to a topological invariant $\nu_{j=0} = -1$~\cite{Kitaev2001}. Consequently, the topological conditions require that the remaining blocks ($j\neq0$) should be topologically trivial.
For the $j\neq0$ blocks, the energy splitting $2\delta^{'}$ between the $j_e$ and $1-j_e$ subbands is given by $\frac{A_1}{R_0}|1-\phi_{\rm flux}|$.
When a magnetic flux of $\phi_{\rm flux} = 1$ is applied, the $j_e$ and $1-j_e$ subbands become perfectly degenerate, indicating a topologically nontrivial system regardless of the Fermi level's position within the bulk gap of the TI nanowire~\cite{Cook-prb-2011,Cook-prb-2012}.
Remarkably, we find that the system always remains topologically nontrivial even when the Fermi level is deep within the conduction band [Fig.~\ref{phase_diagram}(a)]. This finding seems to contrast with previous works that neglected the electrostatic environment and posited that achieving MZMs requires the Fermi level within the bulk gaps~\cite{Cook-prb-2011,Cook-prb-2012,Legg-prb-2021}. To grasp this distinction intuitively, one can apprehend it as the following.
The Fermi levels in the previous works are tuned by the phenomenal parameters, i.e., the homogeneous chemical potential $\mu$. When $\mu$ is inside the conduction band, the bulk of the nanowire becomes metallic and the topological surface states disappear~\cite{Hosur2011,Cook-prb-2011,Cook-prb-2012}. However, in this work, the upward shift of the Fermi level is caused by the band bending effect at the TI-SC interface, described by the electrostatic potential. Although the surface states and bulk states are both occupied, they are confined to an accumulation layer adjacent to the TI-
SC interface. Remarkably, the confinement of the electrostatic potential 
protects the surface states, especially for $j_e = 1/2$ surface subbands, from hybridization with the conduction bands [Fig.~\ref{potential-band}(f)].
This finding is our central result, as it demonstrates that MZMs can be realized even in the presence of conduction bands, which 
are not affected by the band bending effect. Notably, our results require that the primary TI nanowire (without the effect of electrostatic potential) be bulk-insulating, which is consistent with the previous work. 


In addition to the topological invariant $\nu = -1$, the realization of robust MZMs also requires large $\Delta_{\rm{min}}$. Fig.~\ref{phase_diagram}(b) shows the topological phase diagram as a function of the magnetic flux $\phi_{\rm flux}$ and band bending strength $W$. A large topological region with a finite SC gap exits near a single flux quantum, $\phi_{\rm flux} =1$. Notably, 
topological phases are not dependent upon the precise value of $W$. This signifies that achieving MZMs solely demands the application of a magnetic field near the $n = 1$ region, thereby obviating the necessity for finely tuning the Fermi level.
To further confirm the system is exactly in the topological
phase under such conditions, we consider a TI nanowire with finite length $L_z$ in the $z$ direction. Then we calculate the eigenvalues of each $j$ block, as shown in Fig.~\ref{phase_diagram}(c).
Analogous to the Caroli-deGennes-Matricon (CdGM) states~\cite{Caroli1964}, we observe in-gap states with nearly equal energy separation $\delta E$ in each $j$ block. These CdGM analogs are confined
to the TI-SC interface
rather than around a vortex core~\cite{Kopasov-prb-2020,Jose-PRB-2023}. Notably, a pair of MZMs emerge in the $j = 0$ blocks because of the particle-hole symmetry. We further calculate the distribution of the DOS of MZMs in the $L_z-r$ plane [Fig.~\ref{phase_diagram}(d)]. As we can see, MZMs are mostly localized in the center of the top and bottom surface of the TI nanowire and gradually decay to the lateral boundary.
As previously mentioned, the suppression of the SC gap can be mitigated by increasing the radius $R_0$. Nevertheless, the energy separation $\delta E$ diminishes with increasing $R_0$~\cite{Kopasov-prb-2020}, see the black line in Fig.~\ref{phase_diagram}(e).
In order to detect and manipulate
MZMs, it is requisite that $\delta E$ far exceeds the experimental temperature. Notably, $\delta E$ still remains approximately at $0.064 ~\Delta_{0} \approx 0.1$ meV when $R_0 = 100$ nm. 
 Finally, we consider the disorder effect on the TI nanowire.
This is important because present-day bulk insulating TI wires are relatively dirty~\cite{Muenning-NC-2021a}. The charged impurities in the bulk samples can lead to fluctuations of the electrostatic potential up to several meV~\cite{Skinner-prb-2013,Borgwardt-PRB-2016,Knispel-prb-2017,Woods-PRA-2021}. In order to investigate the stability of MZMs, we use the on-site fluctuations in the potential $\delta \phi(r)$  that are drawn randomly as $\delta \phi(r) \in [-u_0/2, u_0/2]$.
Notably, MZMs remains
strongly pinned to zero energy up for $u_0 = 10$ meV [Fig.~\ref{phase_diagram}(f)]. This stability against disorder is a direct consequence of our setup's large topological phase for MZMs.

\section{Discussion and conclusion}\label{sec_conclusion}
This study delves into the topological characteristics of a TI nanowire covered by a full SC shell. To comprehensively account for the system's electrostatic environment, we employ the self-consistent Schr\"odinger-Poisson method, enabling us to compute the internal electrostatic potential within the TI nanowire.
Our analysis unearths a distinctive outcome: the band bending effect at the interface between the TI and SC induces a notable shift of the Fermi level into the conduction band. This shift, in turn, leads to the coexistence of occupied surface states and bulk states, localized within an accumulation region proximate to the TI-SC interface. This accumulation layer maintains a nearly consistent thickness of approximately 30 nm, regardless of the specific radius of the TI nanowire.
When magnetic flux is applied, the surface states and bulk states have different flux-penetration areas, which engenders a suppression on the superconducting gap. To address this issue, we propose to use TI nanowires with larger radii.
Finally, We demonstrate that MZMs can be achieved across a wide spectrum of parameters centered around one applied flux quantum, $\phi_0 = h/2e$. 
Importantly, within this regime, MZMs can be realized even in the presence of conduction bands, which are not affected by the band bending effect.

In our calculations, we have retained the rotational symmetry of the TI nanowire. This strategic choice reduces the computational cost, and facilitates the treatment of the fully three-dimensional system~\cite{Li-NSR-2022}. Importantly, the topological properties of TI nanowires remain insensitive to the specific shape of the cross-section~\cite{Fernando-sci-2019}. 
Refs.~\cite{Vuik-IOP-2016,Antipov2018,Mikkelsen2018} proposed that
the electrostatic environment in Rashba semiconductors has a significant effect on their topological proprieties.  This prompts our inquiry into the electrostatic influences within TI nanowires. Indeed, in the context of TI nanowires, the role of the electrostatic effect also remains essential. 
Compared with bulk states in Rashba semiconductors, the surface states are more localized near the TI-SC interface, so they are more sensitive to band bending effect. Building upon this insight, Ref.~\cite{Chen-prb-2023} demonstrated that surface states near the SC nearly do not respond to gating,  thereby constraining the tunability of the system. As it has been shown theoretically, the geometry of SC in the TI nanowire-based devices can either be a full shell~\cite{Fernando-sci-2019}, or just attaching the SC to several side surfaces of the TI nanowire~\cite{Legg-prb-2021}.  The full shell geometry offers a larger induced SC gap but restricts the tunability of Fermi level through the gate voltage.  Notably,  our results demonstrate that the presence of MZMs remains independent of the band bending strength, thereby eliminating the need for the fine tuning of the Fermi level.  This signifies that achieving MZMs solely demands the application of a magnetic field near the $\phi_{\rm flux}  =h/2e$ region, further reducing the difficulties in experimental control.


\begin{acknowledgements}
{\em Acknowledgments - } Authors thank Chun-Xiao Liu and Fu-Chun Zhang for helpful discussions. X. Liu acknowledges
the support of the Innovation Program for Quantum Science and Technology (Grant No. 2021ZD0302700) and the National Natural Science Foundation of China
(NSFC) (Grant No. 12074133). Dong E. Liu acknowledges the support of the Innovation Program for Quantum Science and Technology (Grant No. 2021ZD0302400), and the National Natural Science Foundation of China (1974198).
Xiao-Hong Pan acknowledges the support of the China Postdoctoral Science Foundation (Grant No. 2023M731208). Zan Cao acknowledges the support of the National Natural Science Foundation of China (No. 12374158).

Li Chen and Xiao-Hong Pan contributed equally to this work.
\end{acknowledgements}

\appendix

\section{Model Hamiltonian of TI in Cylindrical Coordinates}\label{Appendix A}
\setcounter{equation}{0}
\renewcommand\theequation{A\arabic{equation}}

The model Hamiltonian of TI in the Cartesian coordinates takes the form~\cite{Zhang2009_NP,Liu-PRB-2010} 
\begin{equation} \label{Eq:TI}
H_{\textrm{car}}(k) = \epsilon_{0}(k) +
\begin{bmatrix}
M(k) & A_{1}k_z  & 0 & A_{2} k_{-} \\
A_{1}k_z & -M(k) & A_{2} k_{-}  & 0 \\
0  & A_{2} k_{+} & M(k) & A_{1}k_z 
\\
A_{2} k_{+} & 0  & A_{1}k_z & -M(k)
\end{bmatrix},
\end{equation}
where $k_{\pm}=k_{x} \pm i k_{y}$, $\epsilon_{0}(k)=  C_{0} + D_{1}k_{z}^{2} + D_{2}(k_{x}^{2}+k_{y}^{2})$ and $M(k)=M_{0} + B_{1}k_{z}^{2} +B_{2}(k_{x}^{2}+k_{y}^{2})$. To rewrite Eq.~\ref{Eq:TI} in cylindrical coordinates, we can use the relation:
\begin{equation} \label{Eq:transform}
\begin{bmatrix}
\partial_x  \\
\partial_y \\
\partial_z 
\end{bmatrix}  = 
\begin{bmatrix}
\cos{\theta} & -\frac{1}{r}\sin{\theta}  & 0  \\
\sin{\theta}& \frac{1}{r}\cos{\theta}  & 0 \\
0  & 0 & 1
\end{bmatrix}
\begin{bmatrix}
\partial_r  \\
\partial_{\theta} \\
\partial_{z}. 
\end{bmatrix}. 
\end{equation}
Subsequently, the TI Hamiltonian in cylindrical coordinates takes the form
\begin{eqnarray}\label{Eq:TI-sy}
&&H_{\textrm{TI}}(r,\theta,z) = \epsilon_{0}(r,\theta,z) + \\ \nonumber
&&
\begin{bmatrix}
M(r,\theta,z) & -iA_1 \partial_z  & 0 & A_{2} P_{-\theta} \\
-iA_1 \partial_z & -M(r,\theta,z) & A_{2} P_{-\theta}  & 0 \\
0  & A_{2} P_{+\theta} & M(r,\theta,z) & -iA_1 \partial_z
\\
A_{2} P_{+\theta} & 0  & -iA_1 \partial_z & -M(r,\theta,z)
\end{bmatrix}
\end{eqnarray}
where $M(r,\theta,z) = m_0 - B_1 \partial_{z}^{2}-B_2 \nabla_{\textrm{in}}^2$, $\epsilon(r,\theta,z) = C_0 - D_1 \partial_{z}^{2}-D_2 \nabla_{\textrm{in}}^2$. Here, $\nabla_{\textrm{in}}^2 = \frac{1}{r}\partial_r +\partial_{r}^{2}+
\frac{1}{r^2}\partial_{\theta}^2$ is the Laplacian operator in the in-plane coordinates. $P_{\pm \theta} = -i e^{\pm i \theta}(\partial_r \pm \frac{i}{r}\partial_{\theta})$. 

Now, consider a magnetic field applied along the nanowire ($z$ direction), and choose the gauge $\vec{A} = \frac{1}{2}(\vec{B} \times \vec{r}) = A_{\theta} \hat{\theta}$ with $A_{\theta} = \frac{Br}{2}$. It is straightforward to demonstrate that the vector potential affects only $\partial_{\theta}$:
\begin{equation}
    -i\partial_{\theta} \longrightarrow -i\partial_{\theta}-\Phi(r).
\end{equation}
Here, $\Phi(r)=Br^2/\Phi_0$ represents the normalized magnetic flux with respect to the flux quantum $\Phi_0 =h/e$. Subsequently, the TI Hamiltonian changes to:
\begin{equation}
H_{\textrm{TI}} \longrightarrow H_{\textrm{TI}} +H_{M}.
\end{equation}
Where $H_{M}$ is the additional term originating from the magnetic flux, taking the form:
\begin{equation}
H_{M} = \frac{B_2}{r^2}[\Phi^2(r)+2iB_2\Phi(r)\partial_\theta]s_0\sigma_z-\frac{A_2\Phi(r)}{r}s_{\theta}\sigma_x.
\end{equation}
Finally, the TI Hamiltonian with magnetic flux and electrostatic potential in cylindrical coordinates takes the form:
\begin{equation}\label{appendix-Ham_e}
H_{\rm e} = H_{\rm TI} + H_{M}- e\phi(r).
\end{equation}
Notably, we have $[H_{\rm e}, \hat{J}^{\rm e}_{z}] = 0$ with $\hat{J}^{\rm e}_{z} = -i \partial_{\theta}+\frac{1}{2}s_z$. Importantly, the angular dependence of $H_e$ can be eliminated using a unitary transformation $\tilde{H}_{\rm e} = UH_{\rm e}U^{\dagger}$ where $U = \textrm{exp}[-i(j_e-\frac{1}{2}s_z)\theta]$. Consequently, $\tilde{H}_{\rm e}$ becomes block diagonal, expressed as:
\begin{equation}\label{appendix-block-1}
\tilde{H}_{\rm e} = \bigoplus_{j_e,k_z} H_{\rm TI}^{j_{\rm e}}(r,k_z).
\end{equation}
Notably, we replace $-i\partial_z$ with $k_z$ because it is a good quantum number. Then we can divide the $H_{\rm TI}^{j_{\rm e}}(r,k_z)$ into three parts as:
\begin{equation}
H_{\rm TI}^{j_{\rm e}}(r,k_z) = H_{r}^{j_{\rm e}}(r) + H_M^{j_e}(r) + H_{k_z}^{j_{\rm e}}(k_z).
\end{equation}
All the $k_z$ terms are included in $H_{k_z}^{j_{\rm e}}(k_z)$, expressed as 
$ H_{k_z}^{j_{\rm e}}(k_z) = (C_0+D_1 k_z^2)s_0\sigma_0 + (m_0+B_1 k_z^2)s_0\sigma_z+A_1 k_z s_z \sigma_x.$
$H_M^{j_e}(r)$ is the flux term obtained from the transformation $U H_M U^{\dagger}$, which takes the form
\begin{eqnarray}
H_{M}^{j_e}(r) &=& B_2\frac{\Phi^2(r)}{r^2}s_0\sigma_z -\frac{2B_2\Phi(r)}{r^2}(j-\frac{1}{2}s_z)\sigma_z  \nonumber \\ 
&-& \frac{A_2\Phi(r)}{r}s_y\sigma_x. 
\end{eqnarray}\label{eq_ap_H_m}
And $H_{r}^{j_{\rm e}}(r)$ is given by
\begin{eqnarray}\label{Eq:H_r}
&& H_{r}^{j_e}(r) = \epsilon_{r} -e\phi(r)+ \\ \nonumber
&& \begin{bmatrix}
M^{\lambda_{j_e}-\frac{1}{2}} & 0  & 0 & P_{j_e}^{+} \\
0 & -M^{\lambda_{j_e}-\frac{1}{2}} & P_{j_e}^{+}  & 0 \\
0  & P_{j_e}^{-} & M^{\lambda_{j_e}+\frac{1}{2}} & 0
\\
P_{j_e}^{-} & 0  & 0 & -M^{\lambda_{j_e}+\frac{1}{2}}
\end{bmatrix}
\end{eqnarray}
where $\epsilon_r = -D_2(\partial_r^2 +\frac{1}{r}\partial_r-\frac{(\lambda_{j_e} + \frac{1}{2})^2}{r^2} )$, $M^{\lambda_{j_e} \pm \frac{1}{2}} = -B_2(\partial_r^2 +\frac{1}{r}\partial_r-\frac{(\lambda_{j_e}\pm \frac{1}{2})^2}{r^2} )$, $P_{j_e}^{\pm} =  -iA_{2} (\partial_r \pm \frac{\lambda_{j_e} \pm \frac{1}{2}}{r})$. Notably, $\lambda_{j_e} = j_e - \Phi(r)$ which can be regarded as the flux modulated angular momentum. 
Consequently, the blocks with the $\pm \lambda_{j_e}$ have the same eigenvalues.
For instance, when $\Phi(r) = 1/2$, the surface subbands in $j_e = 1/2~(\lambda_{j_e} = 0)$ block are gapless nondegenerate bands. While the subbands within $j_e = -1/2~(\lambda_{j_e} = -1)$ and $j_e = 3/2~(\lambda_{j_e} = 1)$ blocks are degenerate with the same eigenvalues.

\section{Bessel Expansion}\label{Appendix B}
\setcounter{equation}{0}
\renewcommand\theequation{B\arabic{equation}}

\begin{figure}[!htb]
\centerline{\includegraphics[width=1\columnwidth]{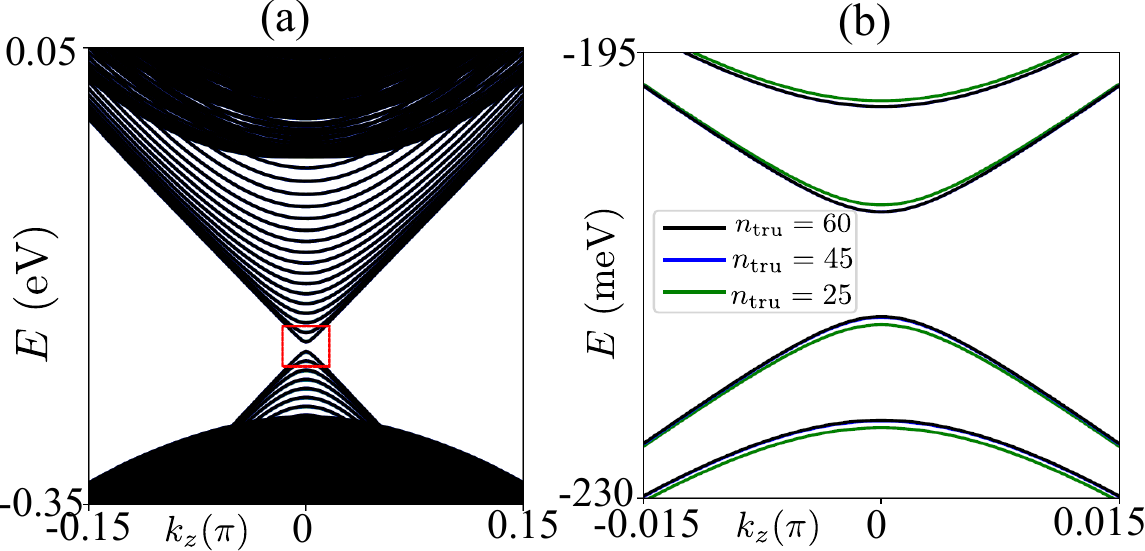}}
\caption{ (a) The electron band structure of TI nanowire. The green, blue, and black lines correspond to the case when $n_{\textrm{tru}}=25,~45,~60$, respectively. (b) Zoom view of the region depicted by the red box in panel (a). It is noted that the band structure converges to stable values when $n_{\textrm{tru}}$ is sufficiently large. In our calculations, we choose $n_{\textrm{tru}} = 45$. Parameters used in this plot: $\phi = 0$, $W = 0.3$, $R = 50$ nm.  }
 \label{fig_N_bessel}. 
\end{figure}

In the main text, we have transformed the Hamiltonian $H_e$ in the block diagonal form according to a unitary transformation $\tilde{H}_{e} = UH_{e}U^{\dagger}$.
Finally, 
$\tilde{H}_{e}$ is  
block diagonal which can be written as 
\beqn\label{block-1}
&&\tilde{H}_{\rm e} = \bigoplus_{j_e,k_z} H_{\rm TI}^{j_{\rm e}}(r,k_z).
\eeqn 
For the numerical diagonalization of the Hamiltonian within each $j_e$ block, we have employed the Bessel expansion. The Bessel functions satisfy the orthogonality relation:
\beqn\label{Bessel_basis}
\frac{1}{(N_{q}^{m})^2}\int_{0}^{R_0}J_{m}(\alpha_{q^{'}}^{m}\frac{r}{R_0})J_{m}(\alpha_{q}^{m}\frac{r}{R_0})rdr = \delta_{qq^{'}},
\eeqn
Here, $m$ denotes the orbital angular momentum and $\alpha_{q}^{m}$ represents the $q$-th zero of the $m$-order Bessel function $J_{m}(x)$. The normalized factor is denoted as $N_{q}^{m} = \frac{1}{\sqrt{2}}R_0J_{m+1}(\alpha_{q}^{m})$. For convenience, we introduce the normalized Bessel functions $\ket{J_{m}^{q}} = J_{m}(\alpha_{q}^{m}\frac{r}{R_0})/N_{q}^{m}$. These normalized functions $\ket{J_{m}^{q}}$ with the same $m$ but different zeros constitute a complete orthogonal basis, suitable for the expansion of the Hamiltonian $H_{\rm TI}^{j_{\rm e}}(r,k_z)$. Because $\ket{J_{m}^{q}}$ have an infinite number of zeros, a truncation is needed. This truncation involves selecting a finite set of zeros, up to a truncated zero $\alpha_{n_{\textrm{tru}}}^{m}$. In this context, the dimension of the discrete Hamiltonian within each $j_e$ block becomes $4\times n_{\textrm{tru}}$. When $n_{\textrm{tru}}$ is chosen sufficiently large, this truncation introduces minimal error within the low-energy regime [Figure~\ref{fig_N_bessel}].

\section{Charge density}\label{Appendix C}
\setcounter{equation}{0}
\renewcommand\theequation{C\arabic{equation}}

\begin{figure}[htb]
\centerline{\includegraphics[width=1\columnwidth]{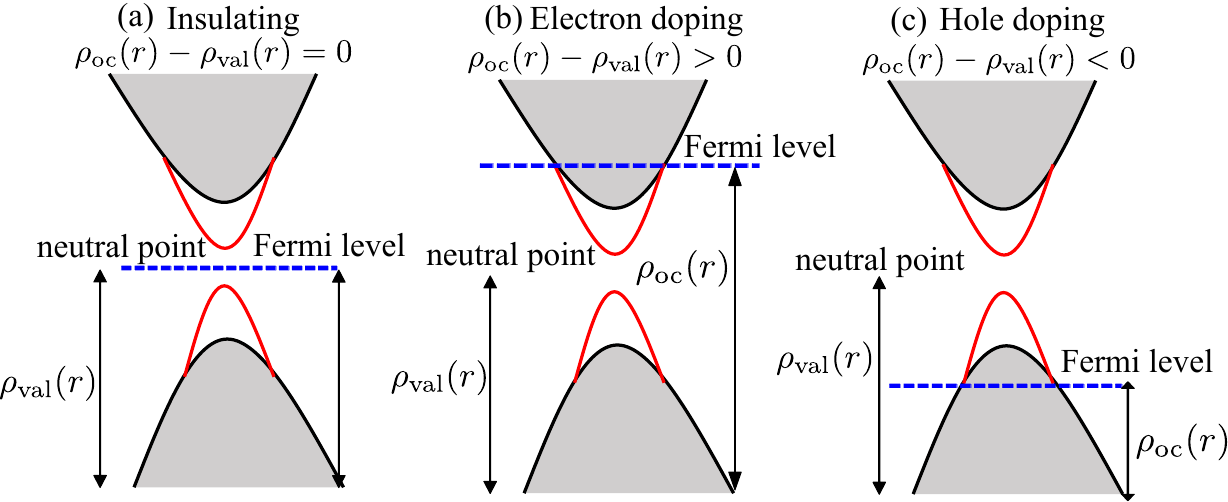}}
\caption{ Schematic diagram of the three typical cases of the band structure of TI nanowire: (a) insulating, (b) electron doping, (c) hole doping. The red and gray bands correspond to TI surface states and bulk states, respectively. The blue dash line represents the Fermi level. $\rho_{\textrm{oc}}(r)$ is defined as the occupied charge density, which is obtained by integrating over the whole occupied
eigenstates. $\rho_{\textrm{val}}(r)$ is the density stems from the whole
valence band. The free electrons or holes is obtained by  $\rho(r) = \rho_{\textrm{oc}}(r) - \rho_{\textrm{val}}(r)$.         }
\label{appendix_fig_rho}
\end{figure}

In our calculations, the Hamiltonian of the topological insulator (TI) is described by a four-band $k\cdot p$ model, which includes both conduction and valence bands. In Fig.~\ref{appendix_fig_rho}, we present a schematic diagram of the TI nanowire's band structure. We define $\rho_{\textrm{oc}}(r)$ as the occupied charge density, obtained by integrating over all occupied eigenstates, while $\rho_{\textrm{val}}(r)$ represents the density originating from the entire valence band. The density of free electrons or holes is given by $\rho(r) = \rho_{\textrm{oc}}(r) - \rho_{\textrm{val}}(r)$.
When the Fermi level is situated at the neutral point, the TI nanowire behaves as an insulator, and we have $\rho_{\textrm{oc}}(r) - \rho_{\textrm{val}}(r) = 0$. However, when the Fermi level is located within the conduction (valence) bands, we have $\rho_{\textrm{oc}}(r) - \rho_{\textrm{val}}(r) > (<)0$, indicating electron (hole) doping. The growth of a TI with SC films induces electron doping from the SC to the TI, causing an upward shift of the Fermi level into the conduction band~\cite{Xu-prl-2015,Ruessmann2022}.

\section{Schr\"odinger-Poisson Method}\label{Appendix D}
\setcounter{equation}{0}
\renewcommand\theequation{D\arabic{equation}}

\begin{figure}[!htb]
\centerline{\includegraphics[width=1\columnwidth]{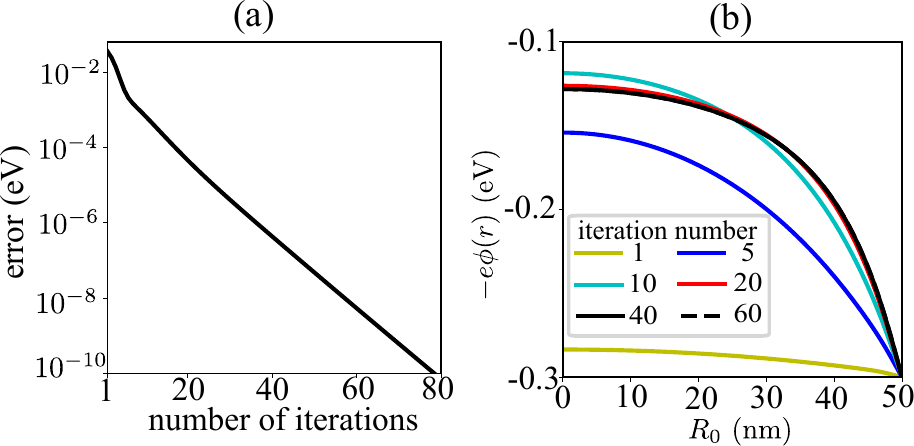}}
\caption{ (a) The error of Schr\"odinger-Poisson equations as a function of the number of iterations. (b) The distribution of the electrostatic energy $-e\phi(r)$ as the number of iterations increases. The convergence occurs when the iterations number $i>40$ 
with the error $\sigma < 10^{-7}$ eV, see the black solid and dashed lines.}
 \label{fig_SP}. 
\end{figure}

 \begin{figure*}[!htb]
\centerline{\includegraphics[width=2\columnwidth]{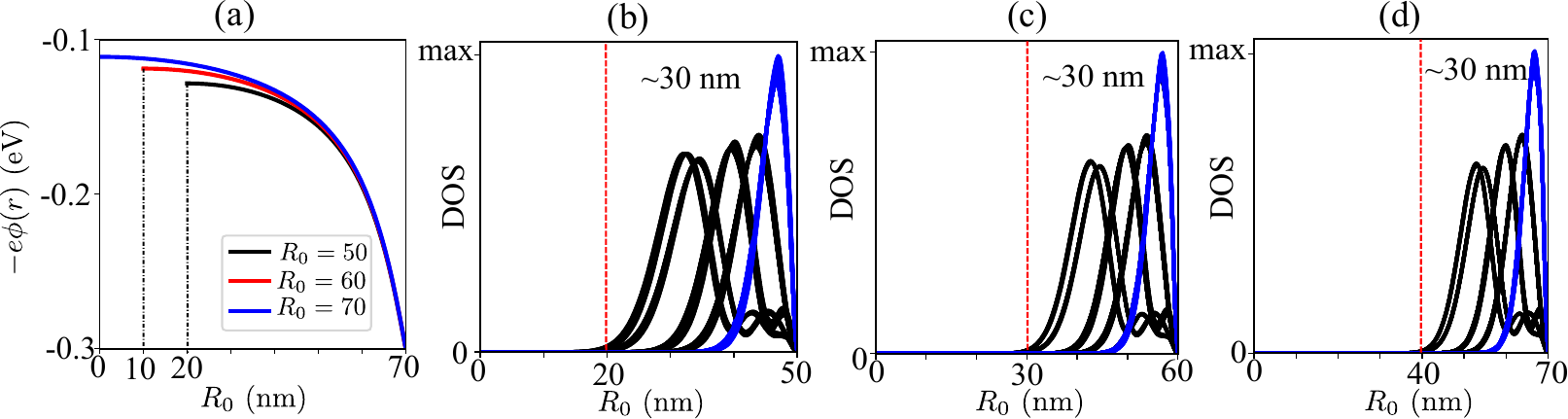}}
\caption{ (a) The distribution of the electrostatic energy $-e\phi(r)$ with different $R_0$. The black, red, and blue lines correspond to $R_0 =50,~60,~70$ nm, respectively. For the convenience of comparison, we align the three different $\phi(r)$ at the boundary of the nanowire, i.e., at the point $r=R_0$. The distribution of the three potentials near the boundary ($r = R_0$) is nearly identical, ensuring a consistent thickness for the accumulation layer. (b)-(d) The density distribution of the occupied states at the Fermi level with  $R_0 =50,~60,~70$ nm, respectively. The blue (black) lines correspond to surface states (bulk states). The accumulation layer has a fixed
thickness of approximately 30 nm, as indicated by the red dashed line.  }
\label{app-aculayer}
\end{figure*}

To obtain the electrostatic potential $\phi(r)$, we employ the Schr\"odinger-Poisson Method. Initially, we introduce an initial potential $\phi_0(r) = 0.1$ eV into the Hamiltonian $H_{\rm TI}^{j_{\rm e}}(r,k_z)$ and solve the Schr\"odinger equation within each $j_e$ block:
\begin{equation}\label{Appendix_b_sp1}
H_{\textrm{TI}}^{j_{\rm e}}(r,k_z) \psi_{n_z,k_z}^{j_e}(r) = E_{n_z,k_z}^{j_e}\psi_{n_z,k_z}^{j_e}(r).
\end{equation}
This yields a set of eigenenergies $E_{n_z,k_z}^{j_e}$ and eigenstates $\psi_{n_z,k_z}^{j_e}(r)$. Here, $n_z$ denotes the index of transverse modes. The charge density $\rho_1$ with potential $\phi_0(r)$ is obtained by integrating over the occupied eigenstates:
\begin{equation}\label{Appendix_b_charge}
\rho_1(r) = \frac{-e}{(2\pi)^2}\sum_{ n,j_e}\int dk_z \left[| \psi_{n_z,k_z}^{j_e}(r) |^2 f_T-\rho_{\textrm{val}}(r)\right].
\end{equation}
Finally, a new potential $\phi_1(r)$ is determined by solving the Poisson equation:
\begin{equation}\label{Appendix_b_sp2}
\frac{1}{r}\partial_r \phi_1(r) + \partial_r^2 \phi_1(r) = - \frac{\rho_1(r)}{\epsilon_0 \epsilon_r}.
\end{equation}

It's worth noting that $\phi_1(r)$ generally deviates from the initial potential $\phi_0(r)$. The discrepancy is quantified by the error:
\begin{equation}\label{Appendix_b_sp3}
\sigma_1 = \frac{\sum_m [\phi_1(r_m)-\phi_0(r_m)]}{N_m}.
\end{equation}
Here, $\sigma_1$ is indexed by the iteration number, $m$ denotes the site index, and $N_m$ is the number of sites. The SP problem necessitates a self-consistent solution involving the iterative equations, Eq.\ref{Appendix_b_sp1} and Eq.\ref{Appendix_b_sp2}, until the error of the $i$-th iteration $\sigma_i$ becomes smaller than the critical value $\sigma_c$. The output $\phi_i(r)$ after convergence is the final self-consistent potential.
In our approach, we utilize the linear iteration. The input potential at each iteration is a mixture of the input and output potentials from the previous iteration~\cite{Mikkelsen2018,Legg-prb-2021}:
\begin{equation}\label{Appendix_b_sp4}
\phi_i^{\textrm{in}}(r) = \kappa \phi_{i-1}^{\textrm{out}}(r)+(1-\kappa)\phi_{i-1}^{\textrm{in}}(r).
\end{equation}
In our calculations, we set $\kappa = 0.1$ and $\sigma_c = 10^{-8}$ eV. The iteration error $\sigma_i$ significantly diminishes as the number of iterations increases [Fig.~\ref{fig_SP}(a)]. The potential convergence is observable after approximately 40 iterations, as illustrated by the black solid and dashed lines [Fig.~\ref{fig_SP}(b)].

\section{Accumulation Layer}\label{Appendix E}
\setcounter{equation}{0}
\renewcommand\theequation{E\arabic{equation}}

The band bending effect-induced electrostatic potential confines the bulk states and surface states to an accumulation layer near the TI-SC interface, with a
characteristic width of approximately 30 nm.  Consequently, the TI nanowire can be approximately divided into two regions: the accumulation layer and the insulating core.
 Remarkably, we find that the  accumulation layer has a fixed
thickness of approximately 30 nm and doesn't increase with
the radius of the TI nanowire. This can be explained by the distribution of the confinement potential [Fig.~\ref{app-aculayer}(a)]. For the convenience of comparison, we align the three different radii $\phi(r)$ at the boundary of the nanowire, i.e., at the point $r=R_0$.  It is evident that the distribution of the three potentials near the boundary ($r = R_0$) is nearly identical, ensuring a consistent thickness for the accumulation layer [Fig.~\ref{app-aculayer}(b)]. Enlarging the nanowire radius will primarily increase the size of the insulating core region. Within this region, due to the absence of charge carriers, the potential remains notably flat.

In a TI-SC hybrid system, a thinner accumulation layer implies a stronger coupling between the TI and the SC, thereby resulting in a more significant proximity effect. Since the thickness of the accumulation layer remains constant irrespective of the radius $R_0$, this property offers an advantage in terms of flexibility in fabricating nanowires under various conditions.
Furthermore, in the presence of magnetic flux, the differing flux-penetration areas between the bulk states and surface states induce a notable reduction in $\Delta_{\rm min}$ [Fig.~\ref{fig_gap_flux}(a)]. By increasing the value of $R_0$, the relative area between the accumulation layer and the nanowire can be effectively reduced. As a consequence, this leads to an enhancement of $\Delta_{\rm min}$ [Fig.~\ref{fig_gap_flux}(b)].

\bibliographystyle{apsrev4-1}
\bibliography{ref}

\end{document}